\documentclass[preprint,article,accept,moreauthors,pdftex]{Definitions/mdpi} 
\usepackage{wrapfig}
\usepackage{wrapfig,lipsum,booktabs}
\firstpage{1} 
\pubvolume{00}
\issuenum{1}
\articlenumber{0}
\pubyear{2022}
\copyrightyear{2022}
\history{Received: date; Accepted: date; Published: date}
\usepackage[normalem]{ulem}
\usepackage{changepage}

\usepackage{subfigure}

\Title{A global sharing mechanism of resources: modeling a crucial step in the fight against pandemics}

\Author{G.K. den Nijs, $^{1}$* %,\dagger}$  
\orcidA{}, J. Edivaldo $^{1}$ %,\ddagger}$
, B.D.L. Ch\^atel $^{2,3}$, J.F. Uleman $^{2,3}$, M. Olde Rikkert $^{2}$, H. Wertheim $^{4}$ and R. Quax $^{1,3}$}

\address{%
$^{1}$ \quad Computational Science Lab, Faculty of Science, University of Amsterdam, Amsterdam, The Netherlands \\%; e-mail@e-mail.com\\
$^{2}$ \quad Department of Geriatric Medicine, Radboudumc Alzheimer Center, Donders Institute for Brain, Cognition and Behaviour, Radboud University Medical Center, %Reinier Postlaan 4, 6525GC, 
Nijmegen, The Netherlands \\
$^{3}$ \quad Institute for Advanced Study, Amsterdam, The Netherlands\\
$^{4}$ \quad Department of Medical Microbiology, Radboud Centre for Infectious Diseases, Radboud University Medical Center, Nijmegen, The Netherlands}

\corres{Correspondence: katinkadennijs@gmail.com}%; Tel.: (optional; include country code; if there are multiple corresponding authors, add author initials) +xx-xxxx-xxx-xxxx (F.L.)}

\abstract{To face pandemics like the one caused by COVID-19, resources such as personal protection equipment (PPE) are needed to reduce the infection rate and protect those in close contact with patients \citep{heymann2020covid, Klompas2021}. The demand for those products increases exponentially as the number of infected grows, outpacing any growth that local production facilities can achieve \citep{ranney2020critical, wu2020facemask}. Disruptions in the global supply chain by closing factories or scaled-down transport routes can further increase resource scarcity\citep{mcmahon2020global}. During the first phase of the COVID-19 pandemic, we witnessed a reflex of `our people first' in many regions, countries, and continents \citep{baldwin2020covid}. In this paper, however, we show that a cooperative sharing mechanism can substantially improve the ability to face epidemics. We present a stylized model in which communities share their resources such that each can receive resources whenever a local epidemic flares up. This can potentially prevent local resource exhaustion and reduce the total number of infected cases. We also show that the success of sharing resources heavily depends on having a sufficiently long delay between the onset of epidemics in different communities. This means that a global sharing mechanism should be paired with measures to slow down the spread of infections from one community to the other. Our work is a first step towards designing a resilient global supply chain mechanism that can deal with future pandemics by design, rather than being subjected to the coincidental and unequal distribution of opportunities per community at present.
}

\keyword{COVID-19; Resilience; Supply Chain; Collaboration; Personal Protection Equipment } % 3 to 10

\begin{document}
%%%%%%%%%%%%%%%%%%%%%%%%%%%%%%%%%%%%%%%%%%

%%%%%%%%%%%%%%%%%%%%%%%%%%%%%%%%%%%%%%%%%%

% \textbf{Significance Statement:} Effective mitigation of a pandemic depends critically on the fast and widespread availability of protective resources, such as face masks and ventilation filters. However, the recent COVID-19 pandemic has demonstrated an `our people first' reflex, increasing the inequality of resource distribution which led to local shortages followed by exacerbation of outbreaks. The naive solution, increasing local production everywhere, is not economical nor feasible as no production facility could ever keep up with the exponential growth of a local epidemic outbreak. In this paper we show instead that a cooperative resource sharing strategy can be effective at mitigating local outbreaks and be potentially beneficial for all parties involved.

\section{Introduction}

To keep a pandemic such as the one caused by COVID-19~\cite{heymann2020covid} under control, resources like test kits and personal protection equipment (PPE; i.e., face masks, gloves, disinfectants) are critical both for the population and for health care personnel \cite{Klompas2021}. Additional resources that have indirect effects on the transmission of an epidemic include having sufficient trained medical personnel, hospital facilities with quarantining capabilities, ventilators, and ventilation. Unfortunately, the demand for resources increases exponentially with the spread of a new epidemic. As the available resource stock is tailored to the population's standard use, this poses serious challenges to the associated resource supply chains. Consequently, several countries with high infection rates reached a state of low to no resource availability during various waves of the COVID-19 pandemic, irrespective of their gross domestic products (GDP)~\cite{ranney2020critical,mcmahon2020global}. 

Due to an exponential increase in demand during epidemic waves, local production increases are rarely fast enough to cover the high need~\cite{wu2020facemask}. Moreover, local production is not the primary provider in many countries which obtain PPE from abroad. In some places, preventive methods were proposed to mitigate the upcoming shortages~\cite{livingston2020sourcing}, such as setting guidelines for rational and safe prolonged PPE use or reuse~\cite{baldwin2020covid} and alternative solutions for protection~\cite{freire2020cotton}. Such mitigation strategies may provide temporary relief but do not address the bigger scalability problem. Moreover, some of these strategies lead to healthcare workers feeling inadequately protected and being fearful of getting infected~\cite{hoernke2021frontline}.

The primary production of medical products comes from global supply chains (GSC). GSCs have been optimized for cost-efficiency and may therefore lack resilience, such as redundant supply lines or large buffers along the chains. For this reason, a major topic of debate is if and how these GSCs should be restructured or replaced~\cite{baldwin2020covid, gereffi2020does, ivanov2020viability}. An example of GSC fragility is the disruption of several technological and health product factories in Wuhan during the COVID-19 pandemic and the closing of shipping ports, impacting the international market and availability of these products on a very short time scale~\cite{ayittey2020economic, singh2020impact}.

Consequently, due to the resource scarcity threat in the first wave of the COVID-19 pandemic, multiple countries started to stock, confiscate, and prohibit the exportation of PPE~\cite{bown2020covid,evenett2020tackling} and diagnostic testing materials. This affected the global market as a whole since no country is fully self-sufficient in all the different types of PPE or test kits. Furthermore, it greatly impacted low-income countries that had to deal with increased prices and the scarcity of goods~\cite{carter2020questions,bown2020covid,espitia2020trade,evenett2020tackling}. Although a plausible solution would be to redesign the GSCs for increased systemic resilience against rare incidents such as pandemics, the restructuring of infrastructure may very well pose a prohibitive cost. Additionally, it would require a globally concerted effort at an unprecedented scale.

Instead, this paper investigates whether implementing a sharing policy among different regions or countries can enhance essential resource stock resilience, just as sharing vaccine doses may increase global viral protection faster \cite{wagner2021vaccine}. A PPE sharing policy differs from increasing the overall resilience of GSCs: instead of ensuring the transport lines from producers to each individual consumer region, a sharing policy entails creating flexible transport lines between the consumer regions. For this reason, it may provide a more feasible approach that is effective at suppressing future pandemics. There are many possible scenarios regarding the duration and developmental forms of COVID-19 epidemics, depending on total adaptive immunity and non-pharmaceutical interventions \cite{SaadRoy2020, haug2020ranking}. It is thus essential to learn from the past and invent feasible strategies now.

To demonstrate its potential effectiveness, we created a stylized model of two communities that can be interpreted as countries, regions, or even cities. We model their local epidemics and the spread of infections across the communities, in combination with their (unspecified) PPE resource. PPE availability lowers the infection rate by offering protection and therefore enables suppressing the infection spread to some degree. We simulated how changes in resource sharing behavior affect the capacity of going through the entire epidemic without fully depleting the PPE stock. Although the computational model is stylized and many more details could be added to enhance faithfulness, the aim of the present paper is to understand at a high level the impact of sharing of essential resources and which factors should be taken into account to reach a better local situation for all communities involved. 

% \textbf{Still cite:} \\
% - \cite{SaadRoy2020}; depending on many factors (mainly factors impacting adaptive immunity, but also nonpharmaceutical interventions), scenarios of if and how (magnitude and timing) long COVID-19 pandemic flare ups still possible.
% - \cite{Klompas2021}; urge that many infections actually happen via presymptomatic patients when incidences are high; thus in healthcare important to not only protect those caring for COVID cases directly, but also patients in general.
% - \cite{haug2020ranking}; suitable combinations of non-pharmaceutical interventions can lead to lower incidence rates. Smart to learn from earlier responses in the pandemic to predict effectiveness of measures to future upflares.
% - \cite{wagner2021vaccine}; vaccine sharing for better control COVID. Thus; similar principle but for vaccines instead of PPE. (Reviewer-option Bryan Grenfell) \\

%%%%%%%%%%%%%%%%%%%%%%%%%%%%%%%%%%%%%%%%%%

\section{Methods}

Here follows a functional description of the model needed to understand the results. Readers interested in more technical details are referred to the Extended Methods in Appendix~\ref{extended_methods}. 

\subsection{Coupled SIR and DSP model}

We developed a simplified Demand-Stock-Production (DSP) model to emulate the usage and production of PPE. The DSP model is coupled to a classical Susceptible-Infected-Recovered (SIR) model~\cite{shulgin1998pulse,allen1994some} to cover the disease transmission dynamics. In this way, we can simulate the increased transmission of a virus due to PPE scarcity. PPE can be interpreted here as any physical equipment that is capable of limiting transmission.  N.B.: throughout this paper the letter $S$ will be used to refer to the PPE `stock' from the DSP model, as susceptibles (from the SIR model) are never referred to outside of the appendix. 

The coupling is arranged via two connecting factors: (1) a community's current PPE demand in the DSP chain depends on the number of infected people in that community; and (2) the effective reproduction number $R_e$ of the SIR model depends on the available PPE stock. The two couplings are based on the assumptions that (1) more protective equipment is needed when the amount of infected increases \cite{ranney2020critical, hoernke2021frontline}; and (2) that the unavailability of PPE increases the infection rate of the disease \cite{mcmahon2020global, wu2020facemask}.

In all experiments, two isolated communities $A$ and $B$ of $n$ individuals each are simulated. They differ only in maximum stock ($S^{\textrm{max}}$) and epidemic onset time. The results are described from the point of view of community $A$. Therefore, epidemic onset in $A$ is always at $t=0$, with the introduction of one infected individual into this community. The epidemic onset in $B$ is at a certain time difference $\Delta t$ from $t=0$, thus at $t=\Delta t$. This time difference models the time it would take for the epidemic to spread between the two communities directly, or via another (`confounding') community. The difference can be $\Delta t=0$ days for simultaneous onset and take on both positive and negative values. In this way, both an epidemic starting in $A$ followed by a spread to $B$ and vice versa can be investigated. This enables us to study the effects of changes in the time difference of the epidemic onset between $A$ and $B$ and in their relative order. The differences in maximum stock reflect the different capacities to directly access the necessary PPE, due to e.g. differences in Gross Domestic Product (GDP) and storage capacity. By using different maximum stock levels we can investigate how possible asymmetries between communities might impact sharing behavior outcomes.

\subsection{Sharing system}
We further implemented a PPE sharing system between the two communities to simulate sharing behavior. Sharing is managed via a switch mechanism that depends on the ratio between the current and maximum stock of each community $S/S^{\textrm{max}}$ and a sharing threshold $\theta$. The sharing threshold is a critical stock ratio agreed upon beforehand by the communities. The communities want to ensure always having at least this proportion of stock available to themselves, to increase their chances of weathering a local epidemic without reaching stock depletion ($S=0$). If during an epidemic the stock of a community falls below the sharing threshold $\theta$, they request support from the other community. The switch mechanism activates the sharing system if the other community has the possibility to share, i.e. if their own stock ratio is above the sharing threshold. If the second community starts sharing, they continue until reaching the sharing threshold $\theta$ themselves or until the other community's stock is restored to at least $\theta$ again. In summary, the switch mechanism leads to active PPE sharing if and only if exactly one of the communities' stock levels is below the sharing threshold and the other community has a stock level above the threshold. A low sharing threshold $\theta$ corresponds to communities being willing to share a large fraction of their stock with one another, while a high $\theta$ indicates that they intend to keep a large fraction of stock for themselves.

%%%%%%%%%%%%%%%%%%%%%%%%%%%%%%%%%%%%%%%%%%

\begin{figure}[t!]
    \centering  
    \includegraphics[width=1\textwidth]{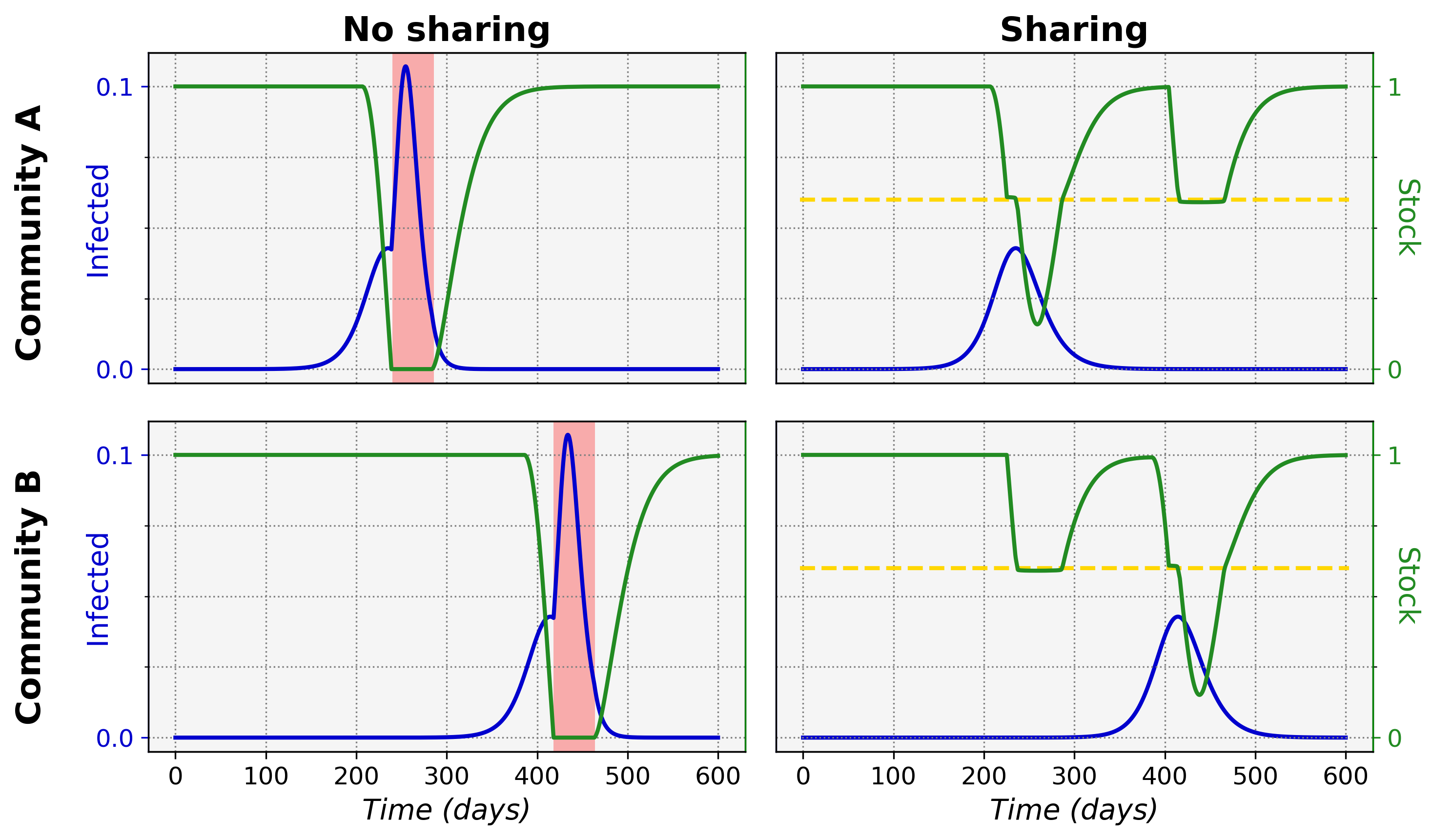} 
    \caption{\textbf{Time evolution of epidemic development in two communities without and with sharing mechanism.}
    Shown are the proportions of infected (blue) relative to the total community population and the proportions of stock (green) relative to the maximum stock over time. The communities had an equal maximum stock of $S^{\textrm{max}}=3\times10^7$. The only difference between the communities was epidemic onset, with the epidemic being introduced in $A$ at $t=0$ and in $B$ at $t=\Delta t=180$ days. Critical zero-stock periods are shaded in red in the no sharing condition (left column). The lack of stock leads to an increase in epidemic spread and a high infected peak. The sharing threshold $\theta=0.6$ (yellow dotted line) is indicated in the sharing condition (right column). Here sharing prevents the total depletion of stock and the associated high infected peak: both the local and total infected ratios are low. $n=1.7\times10^7$, $R_{0}=2.3$, $\gamma=1/6$, $r=0.4$, $w=4$, $P(0)=5\times10^7$, $P^{\textrm{max}}=4\times P(0)$, $D(0)=P(0)$.}
    \label{fig:TimeEvolution}
\end{figure}

\section{Results}
% We distinguished between two broad scenarios: 1) sharing between communities with equal maximum stocks $S^{\textrm{max}}$ and 2) sharing between communities with different $S^{\textrm{max}}$. 
The effects of differences in the relative time difference between communities' epidemic onset, maximum stock height and sharing threshold were assessed in different scenarios that are outlined below. We will first discuss the case where the collaborating communities have equal maximum stocks $S^{\textrm{max}}$ and then the case where their maximum stocks are unequal.

Our main results are shown in the form of phase plots with a distinctive color scheme. We discriminate between three qualitative outcomes in the phase plots, shown in red, white and blue, in decreasing order of severity. In all phase plots, the total infection incidence is given as the ratio of individuals within the populations that has been infected during the disease spread: the infected ratio. If the stocks of both communities would deplete fully at any point during the epidemic, then their effective infection reproduction numbers would increase enormously. This would lead to the highest possible total infected ratio, corresponding to the red state. If only one community would reach zero stock then this would lead to an intermediate total infected ratio, corresponding to the white intermediate state. Finally, if both communities could weather the epidemic without reaching zero stock, the blue state would be reached with the lowest possible infected ratio. Intermediate transition phases, e.g. visible in orange, could appear due to small differences in the total infected ratio, for example due to slight differences in zero stock period duration. 

\subsection{Equal maximum stocks}

We can observe that sharing between certain equal maximum stock communities can prevent stock depletion and the associated high infected ratio. Figure~\ref{fig:TimeEvolution} shows typical epidemic time evolutions for two equal maximum stock communities and the difference between non-sharing and sharing. This clearly depicts how non-sharing (left column) for these communities leads to stock depletion. During a stock depletion period (indicated by the red shading) $R_e$ becomes larger and therefore leads to stronger exponential growth and a high peak of infections. In the sharing condition (right column) we can see that stock sharing with the community in need can prevent full stock depletion and thus prevent the $R_e$ increase, keeping the associated incidence peak to a manageable level. This ultimately leads to a significantly lower total of infected cases in both communities (area under the infected curves). 

\begin{figure}[t!]
    \centering
    \includegraphics[width=1.05\textwidth]{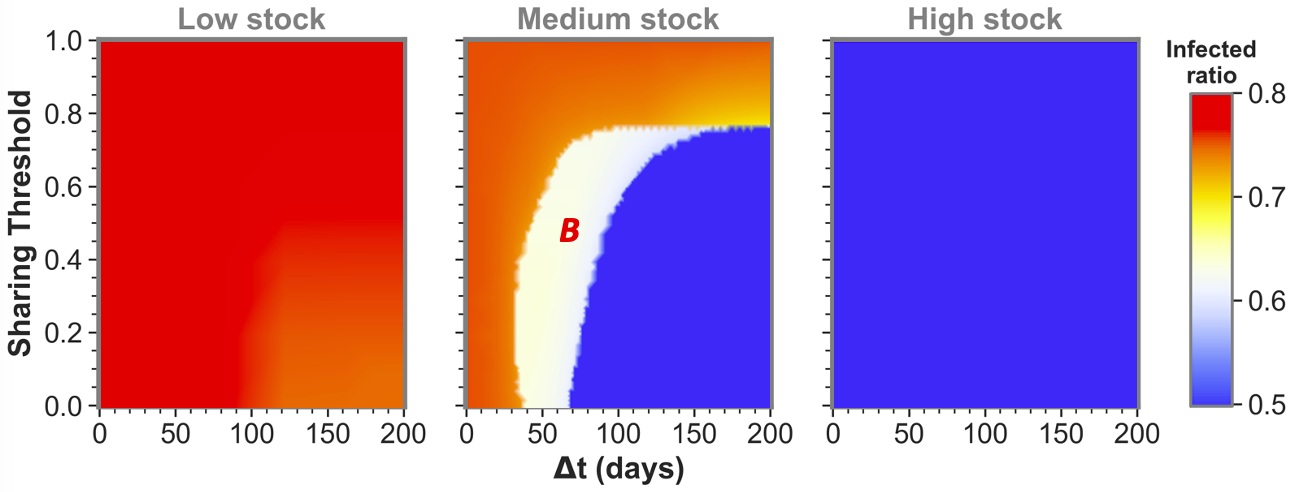}
    \caption{\textbf{Influence of epidemic onset time difference and sharing threshold on total infected ratio for equal maximum stock communities.} The epidemic is introduced in $A$ at $t=0$, and in $B$ at $t=\Delta t$. The indicated infected ratio is the average of the two communities' infected ratios. The red state indicates both communities reaching a high infected ratio, white means that one community (labelled with the red letter, here $B$) had a high infected ratio while the other remained low. In the blue state both communities had a relatively low amount of infected. For each phase plot, the two communities had the same maximum stock, being either $S^{\textrm{max}}=1\times10^7$ (Low Stock), $S^{\textrm{max}}=3\times10^7$ (Medium Stock) or $S^{\textrm{max}}=7\times10^7$ (High Stock). $n=1.7\times10^7$, $R_{0}=2.3$, $\gamma=1/6$, $r=0.4$, $w=4$, $P(0)=5\times10^7$, $P^{\textrm{max}}=4\times P(0)$, $D(0)=P(0)$.}
    \label{fig:Threshold_Delay}
\end{figure}

% Communities as the ones shown here cannot prevent high infected ratios in isolation, but may reach low ratios for one or even both communities when sharing with a low enough threshold and, crucially, a big enough time difference between epidemic onsets. 
The size of the time difference ($\Delta t$) between epidemic onsets heavily influences if communities as the ones shown here can reach low infected ratios for one or even both communities by sharing their stock. This is illustrated by the middle pane of figure~\ref{fig:Threshold_Delay} for medium stock communities (defined to have a maximum stock of $S^{\textrm{max}}=3\times10^7$). When not sharing ($\theta=1.0)$ these communities end up in the red state, both having the highest infected ratio. However, if the sharing threshold is low enough and there is some time difference between the epidemic onsets, the communities can reach the white state in which only one of the two communities ($B$ in this case) reaches stock depletion. Although this is already a successful sharing example for $A$, the time difference is still too short for the combined stocks to be restored sufficiently for the second epidemic in $B$. That being said, we can see that when the time difference between $A$ and $B$ epidemic onsets is increased further, both communities can achieve a low infected ratio (blue state). In the latter case, there was enough time to replenish the stock levels between the consecutive epidemics to save both communities. 

We defined the medium stock category with this maximum stock of $S^{\textrm{max}}=3\times10^7$ to explicitly show the above described behavior when cooperating with a community that is equal in maximum stock. The other distinctive categories we use throughout this paper are low stock ($S^{\textrm{max}}=1\times10^7$) and high stock ($S^{\textrm{max}}=7\times10^7$) communities. Low stock is set at this specific level as these communities show high infected ratios (red state) individually and for all possible conditions when sharing with another low stock group, as can be seen in the left pane of figure~\ref{fig:Threshold_Delay}. Conversely, high stock communities always reach low infected ratios (blue state in the right pane), irrespective of their sharing behavior with an equal stock community. As mentioned before, virtually no actual countries exist where (all possible) PPE stock levels can be high enough to endure a whole epidemic of this sort. The high stock category is therefore included merely for illustrational purposes.

\begin{figure}[t!]
    \centering
    \includegraphics[width=1.05\textwidth]{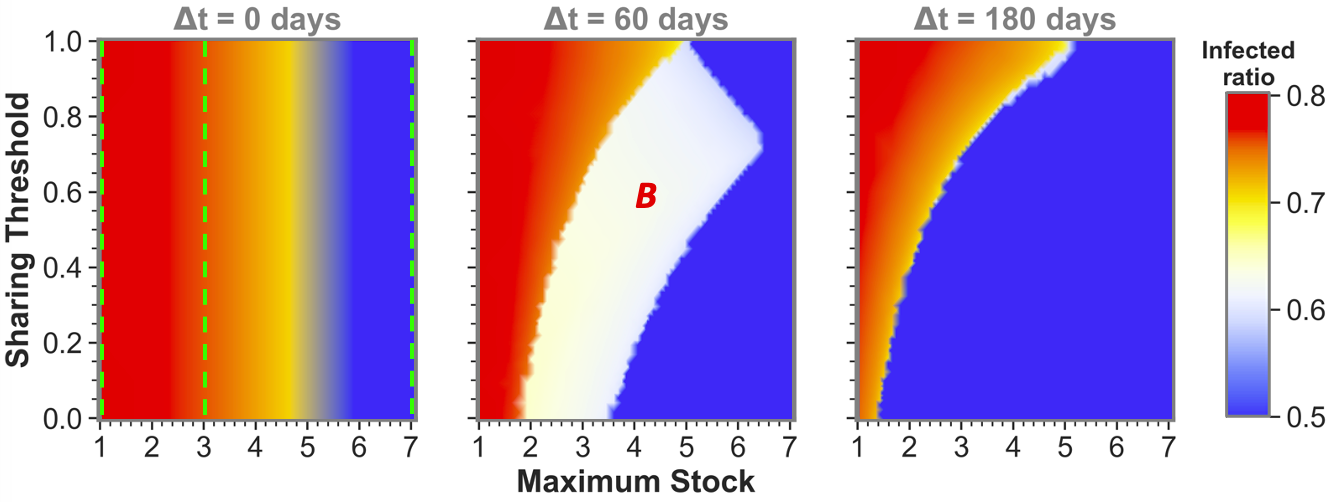}
    \caption{\textbf{Influence of maximum stock size and sharing threshold on the total infected ratio for three distinct epidemic onset time differences between equal maximum stock communities.} In each simulation, the two communities had equal maximum stock sizes. The white state is labelled with the community that reached zero stock and a high infected ratio ($B$). The green dashed lines in the left panel indicate where cross-sections are taken for the three maximum stock categories used throughout this paper: Low Stock with $S^{\textrm{max}}=1\times10^7$, Medium Stock with $S^{\textrm{max}}=3\times10^7$ and High Stock with $S^{\textrm{max}}=7\times10^7$. The values at these lines correspond to those at the different maximum stock phaseplots in figure \ref{fig:Threshold_Delay} for $\Delta t = 0$. Within the $\Delta t=60$ and $\Delta t =180$ phase plots, cross-sections at the same stock values can be imagined for likewise comparison with the previous figure. $n=1.7\times10^7$, $R_{0}=2.3$, $\gamma=1/6$, $r=0.4$, $w=4$, $P(0)=5\times10^7$, $P^{\textrm{max}}=4\times P(0)$, $D(0)=P(0)$.}
    \label{fig:Threshold_MaxStock}
\end{figure}

The observation that sufficient time difference increases the chance of having one or both sharing communities reach a low infected ratio, also holds for more intermediate maximum stock communities. This becomes apparent from figure~\ref{fig:Threshold_MaxStock}, which shows sharing between more intermediate maximum stock sizes for three time difference categories that can be seen as cross-sections (at $\Delta t=0$, $\Delta t = 60$ and $\Delta t =180$) of the phaseplots in figure~\ref{fig:Threshold_Delay} and vice versa (at the maximum stock categories, see green dashed lines). We can observe that equal stock communities can already benefit from sharing from a maximum stock as low as $S^{\textrm{max}}\approx2\times10^7$ for big enough time differences. We also see that the lower the maximum stock, the more important implementing a low sharing threshold is for sharing success. The white area in which community $B$ (the community that gets infected secondly) still reaches high infected numbers even diminishes for the lower maximum stock communities when $\Delta t=180$. 

In summary, the most important result for equal maximum stock community sharing is that communities with insufficient stock to weather the epidemic by themselves can, when agreeing on sufficiently low sharing thresholds, pass the epidemic without reaching the maximal infected ratio. Crucially, for these communities, a large-scale epidemic can be averted only for either or both of them when the time difference between epidemic onsets is sufficiently long. This requires a tailored strategy depending on the dynamics of the disease at hand, but always with a strong focus on mitigating disease spread between different regions by for example imposing travel restrictions early in epidemic outbreaks. 

{
\parfillskip=0pt
\parskip=0pt
% \par}\begin{wrapfigure}{r}{0.50\textwidth}
\par}\begin{wrapfigure}{r}{0.45\textwidth}
  \begin{center}
  \begin{subfigure}
    \centering
    \includegraphics[width=0.45\textwidth]{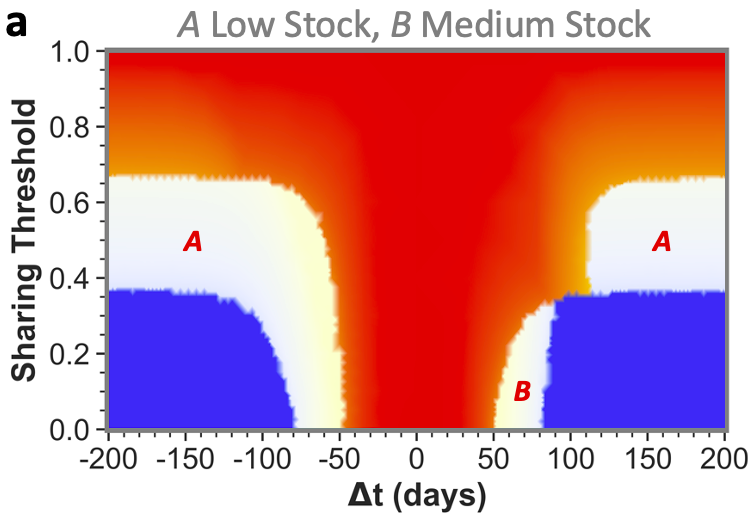}
  \end{subfigure}
  \begin{subfigure}
    \centering
    \includegraphics[width=0.45\textwidth]{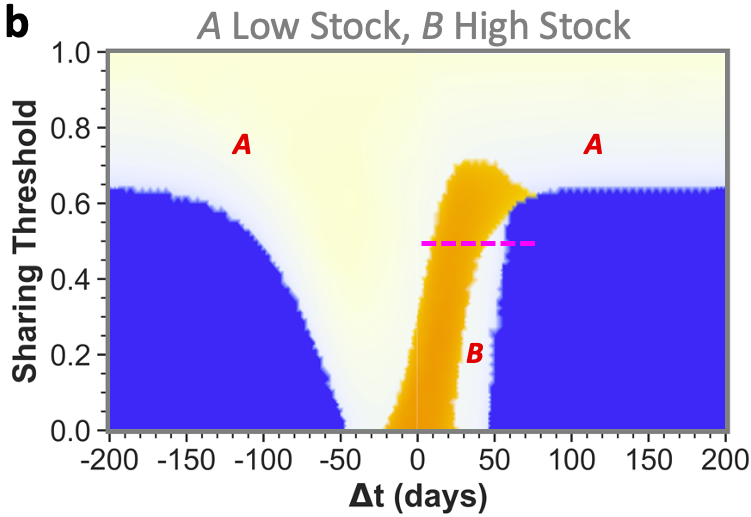}
  \end{subfigure}
  \begin{subfigure}
    \centering
    \includegraphics[width=0.45\textwidth]{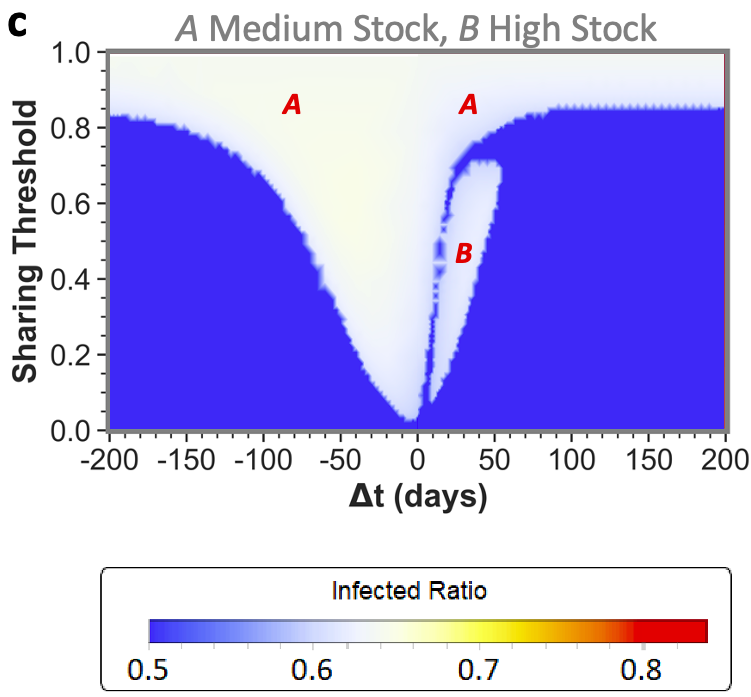}
  \end{subfigure}
  \end{center}
  \caption{\textbf{Influence of time difference between epidemic onset in $\mathbf{A}$ and $\mathbf{B}$ and sharing threshold on total infected ratio for unequal maximum stock communities.} The epidemic is introduced in $A$ at $t=0$, and in $B$ at $t=\Delta t$. In the white areas, a red letter indicates which of the two communities reached zero stock. The pink dashed line in (b) indicates the settings taken for the time evolutions shown in figure~\ref{fig:Time_evolution_2}. $n=1.7\times10^7$, $R_{0}=2.3$, $\gamma=1/6$, $r=0.4$, $w=4$, $P(0)=5\times10^7$, $P^{\textrm{max}}=4\times P_i(0)$, $D(0)=P(0)$.}
    \label{fig:diffMaxStock_C2Delay}
\end{wrapfigure}
\parfillskip=0pt
\parskip=0pt \noindent 

\subsection{Unequal maximum stocks}
Upon reading the equal maximum stock results, one could start fearing that sharing poses a risk to the second community. However, as we will further show by presenting the unequal maximum stock sharing results, this fear is unjustified because (1) for a low or medium stock community, sharing never leads to worse results than non-sharing for both equal and unequal maximum stock combinations and (2) for higher stock communities there is hardly any risk in unequal sharing: even when the total shared stock turns out to be insufficiently high for both communities, a higher stock second community can still weather the epidemic with a low infected ratio (figure~\ref{fig:diffMaxStock_C2Delay}). Only for very specific low time differences between epidemic onsets, a secondly affected higher stock community could possibly be worse off; i.e. experience so-called `second community disadvantage' (in the white areas indicated with $B$ in the figure). This should rather again stress the importance of implementing measures to sufficiently slow down infection spread between communities, than discourage higher stock communities from sharing. Furthermore, in figure~\ref{fig:diffMaxStock_C2Delay}c, we can even observe that a low enough sharing threshold fully abolished the chance for high stock communities of having the second community disadvantage. However, as this was only the case for these very specific disease dynamics and stock levels, communities should not solely rely on a low sharing threshold and still see increasing the onset time difference as top priority. The substantial reduction of the second community disadvantage for higher stock communities is a major difference from what we have seen for equal community sharing, in which the white state always meant full depletion for the community affected secondly by the epidemic. 

An additional interesting observation is that sharing between unequal communities could help both communities weather the epidemic for all maximum stock height combinations. Even the low and medium stock communities can together end up in blue states by sharing! To reach blue states, the ordering of epidemic onsets is not of major importance, as we can see from the approximate similar sizes of the blue states for both positive and negative time differences. What is of major importance to obtain these results, is (again) the time differences between these onsets and agreeing on a low sharing threshold.

In other aspects however, the epidemic onset ordering does impact sharing outcomes, which is indicated by some asymmetries that can be observed in the phaseplots around $\Delta t=0$: an epidemic that starts in the highest stock community (negative $\Delta t$'s) will in general less often lead to the highest total infected ratios (red and orange states) than vice versa. It should be noted, however, that the lowest infected ratio regions (blue states) are often bigger when the lower stock community is infected first (positive $\Delta t$'s). If starting in the lower stock community, we see that increasing the time difference between onsets is of utmost importance to avoid high total infected ratios. The longer this difference between the epidemics, the more time for stock regeneration. 

\begin{figure}[t!]
    \centering
    \includegraphics[width=\textwidth]{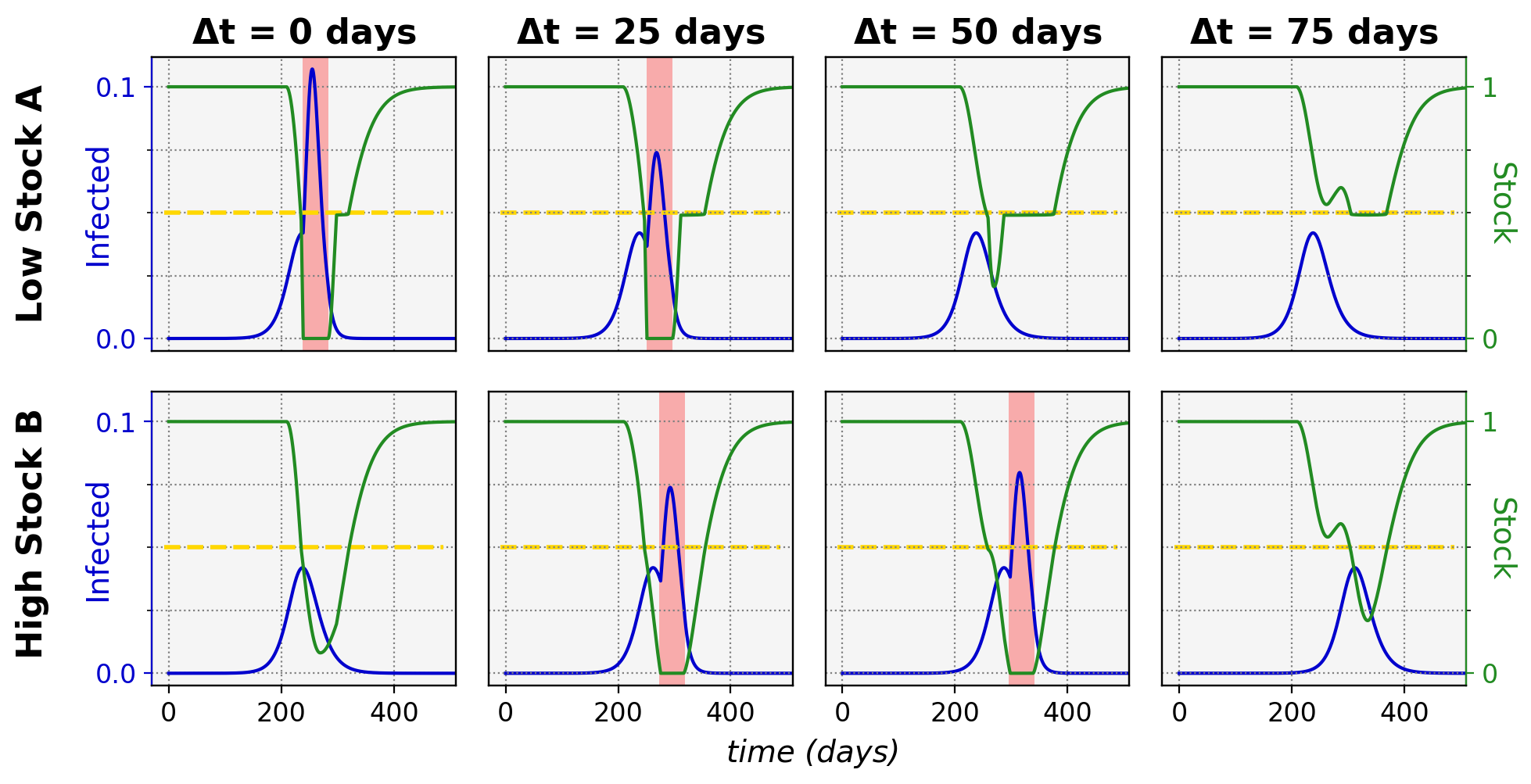}
    \caption{\textbf{Effect of increasing epidemic onset time difference on sharing success of unequal communities.} Indicated are the proportions of infected (blue) relative to the total community population and the proportions of stock (green) relative to the maximum stock over time. Critical zero-stock periods are shaded in red and the sharing threshold ($\theta=0.5$) is indicated by the yellow dotted line. The communities had unequal maximum stocks, with $S^{\textrm{max}}_A=1\times10^7$ and $S^{\textrm{max}}_B=7\times10^7$. By increasing the epidemic onset time difference, the system transitions from only high-stock community $B$ weathering the epidemic with low infected ratios (for $\Delta t=0$ days), via both communities reaching high infected ratios (for $\Delta t=25$ days) and only low stock community $A$ reaching a low infected ratio (for $\Delta t=50$ days), to both having low infected ratios (for $\Delta t=75$ days). This behavior is observed for the specific settings as indicated by the pink dashed line in figure~\ref{fig:diffMaxStock_C2Delay}, but more generally illustrates the impact of epidemic onset time difference on the possible system outcomes. $n=1.7\times10^7$, $R_{0}=2.3$, $\gamma=1/6$, $r=0.4$, $w=4$, $P(0)=5\times10^7$, $P^{\textrm{max}}=4\times P(0)$, $D(0)=P(0)$.}
    \label{fig:Time_evolution_2}
\end{figure}

An interesting illustration of the effects of increasing the time difference between onset in a lower stock and higher stock community is shown in figure~\ref{fig:Time_evolution_2}, for which the relevant sharing threshold and time difference settings are indicated with the pink dashed line in figure~\ref{fig:diffMaxStock_C2Delay}. By increasing the threshold, the system passes from a state of depleted stock for the first community, via the maximal total infected ratio and a state of second community disadvantage, to both communities reaching low infected ratios. This effect is indeed asymmetric around $\Delta t=0$: if time evolutions would be taken for the same threshold, but with negative $\Delta t$'s (the mirror-image of the pink dashed line), we would only observe the state in which the lower stock community depletes fully. We can further observe that the asymmetry in the phaseplots vanishes in the larger time difference regions. This is because, after sufficient time, the first epidemic dynamics die out and the stocks replenish enough, resulting in virtually no interaction effects between the consecutive epidemics.

The results discussed here solely focus on the three stock categories as we defined before to ease result interpretation. For more intermediate stock values, we refer to supplementary figure~\ref{fig:diffMaxStock_C2Init}, where the effect of an increasing $S^{\textrm{max}}_B$ on infected ratio is shown for different fixed $S^{\textrm{max}}_A$ and different onset time difference categories.

Summarizing, the unequal maximum stock sharing results presented here again stress the positive combined effect of a low sharing threshold and sufficient time difference between epidemic onsets. Additionally they show (1) that there is no downside from sharing for lower stock communities and (2) how high stock communities may substantially improve the overall epidemic resilience without much increased risk for the own population, as long as sufficient measures are implemented to reduce epidemic spread between the communities.

\section{Discussion}

We have introduced a stylized PPE production model with a sharing mechanism for communities facing an epidemic. The model convincingly shows that involved PPE stock sharing can greatly improve total and local epidemic resilience for all combinations of maximum stock communities. Crucially, these sharing strategies should be accompanied by disease spread mitigation measures to increase the time between epidemic peaks. This ensures that both communities have access to large quantities of stock when they need it most. First, we saw that completely blocking the circulation of essential supplies can lead to local stock shortages and subsequently epidemic proliferation. Then we showed that for all realistically possible scenarios (where communities never have enough stock to fully endure exceptional situation like the COVID-19 pandemic by themselves), sharing leads to better overall and local epidemic outcomes than non-sharing. Even if a high stock community would exist, sharing can never lead to a more negative outcome for any cooperating community as long as the time differences are long enough. It is good to additionally realise that investing in global reduction of infection rates helps to restrict the total viral circulation and chance of reinfection in each region. Therefore, these investments are beneficial to all and crucial for final pandemic mitigation. The results thus strongly advocate for a willingness to share big stock proportions, with a special focus on accompanying measures to slow down epidemic spread between communities. In this way, pandemics can be fought way more efficiently and successfully for all parties involved with their respective PPE production limitations.

% met gedoe vaccins laat zien dat het geen zin heeft om voor je zelf te gaan, als het ergens blijft circuleren blijf je last hebben. dus je moet anderen helpen,. goed om specifiek te noemen en ook parallel met vaccins

We emphasize that the numbers used in the examples are not intended to be realistic and merely for illustrational purposes. In particular, what time difference would be `sufficient' will depend on the properties of the virus, the modes in which people have contacts with other people, the infrastructure and production capabilities of the PPE, the preventive measures in place, and so on.

As mentioned before: increasing the delay between epidemic onset should evidently be a major focus when implementing this sharing mechanism. To achieve this, participating countries must directly focus on e.g. limiting international travel when a community gets infected. Interestingly, the process of resource sharing itself could then start posing a threat, by increasing the chance of importing infection cases~\cite{killeen2020lockdown}. We believe, however, that setting clear safety guidelines for this process can lower the risk sufficiently to still make sharing the most viable option. For interested policymakers, the model could be extended with the possibility to import infection cases during sharing. In this way this possible risk and the necessary precautions could be investigated in more detail.

There are some simplifications in our model, which could possibly be accounted for in further research. For example, the effective reproduction number $R_e$ only changes when the community's stock fully depletes ($S=0$). In a more realistic scenario, there would be a gradual increase in $R_e$ as the stock becomes more scarce and less people can be protected. This would result in smoother phase plots, abolishing the three clear outcome states observed in this paper. Nevertheless, the general result that sharing reduces the infected ratio will then still hold. This more realistic approach would therefore make the model interpretations more complex without adding new insights. Further, parameters like the basic reproduction number $R_0$, but also the population size $n$ and PPE production rate $P$ could differ substantially between communities. At the moment, the only characteristic we varied between communities was their maximum stock $S^{\textrm{max}}$, but it could be interesting to study the effects of greater heterogeneity between communities on the results. One last simplification worth mentioning is the instantaneous transfer of huge PPE stock quantities during sharing. Although it was beyond the scope of this paper, it could be interesting to extend the model with some geographical transportation limits and delays for different distances to ascertain the boundary conditions for sharing to be effective.

Further interesting extensions could be to include more communities and tailoring the DSP model to specific and possibly multiple resource types. When multiple communities are involved, the necessary individual sharing amount to sufficiently help a community in need reduces as the load can be divided over contributing groups. In our model we have seen that two low stock communities had too little combined stock to save either of them, even when sharing major proportions of stock. But a bigger group of solely low stock communities could potentially have enough combined stock to successfully avoid high infected ratios. However, one should note that in this more complex scenario also a more complex sharing strategy decision structure arises. To model the politics of this, game theoretic approaches could be used~\cite{bianchi2020solidarity,caparros2020corona}. For the extensions suggested here, the modeller should further be aware of the increasing computational costs of solving $3(n+r+1)$ ODE's for $n$ communities and $r$ resource types. Lastly, in all its simplicity, our model could serve as a basis for models of higher complexity that more closely resemble real-world scenarios. After fitting data of manufacturers, resource usage and epidemic dynamics, optimal sharing strategies could potentially be estimated with some precision for combinations of collaborating communities, depending on the uncertainties involved. A similar work of fitting is done to predict SARS-Cov 2 evolution~\cite{neves2020predicting}. 

Overall, we can see that increasing global resilience to future pandemics is a complex problem that requires an interdisciplinary effort. Although myriad extensions could be made to this model, these additions would have unncessarily distracted the reader from our main aim of communicating to policymakers how cooperation can improve epidemic resilience. The strengths of this model are its simplicity and comprehensibility; conveying a clear message and recommendations.

\section{Conclusion}

This work demonstrates how trade barriers for essential healthcare resources can worsen pandemic outcomes, while cooperation between different communities can successfully avoid resource scarcity and prevent high infected incidences for all parties involved. We used a stylized model encompassing disease spread and a resource supply chain in two communities. If taken enough measures to slow down infection spread between them, both communities were able to avoid resource exhaustion at any point during their epidemics by cooperative sharing. This shows that collaboration is a viable solution to avoid resource scarcity. This message is similarly true for vaccines as we have learned during the current COVID-19 pandemic \cite{wagner2021vaccine}.

The model serves as a proof-of-concept that illustrates how sharing protective equipment resources between communities is a crucial step in epidemic mitigation. Although the model is stylized, we believe that its general insights are easy to understand and essential to disseminate to all stakeholders involved. It advocates for involved sharing strategies between all types of communities, showing that with a high willingness to share and a strong focus on slowing international infection spread, total epidemic mitigation outcomes can be improved greatly. It also successfully demonstrates how high GDP countries should not be afraid to offer help to countries that need it, as potential negative effects for their own population can be abolished if the right disease spread mitigation measures are in place. Furthermore, reducing viral circulation globally reduces each region's chance of reinfection and is thus in the interest of all. By taking these insights into account, we believe that global initiatives can help prevent essential resource exhaustion during pandemics. Mutual sharing can potentially support all participating communities and ultimately lead to collaborative mitigation of a pandemic.

%%%%%%%%%%%%%%%%%%%%%%%%%%%%%%%%%%%%%%%%%%
\vspace{6pt} 

\acknowledgments{
%{Acknowledgments}
RQ, MOR, HW acknowledge the ZonMw project ‘Dynamic symptoms networks’ (project number 09120012010063), as well as the contributions made by prof. dr. A. G. Hoekstra and A. Titton.}

%{In this section you can acknowledge any support given which is not covered by the author contribution or funding sections. This may include administrative and technical support, or donations in kind (e.g., materials used for experiments).}

%%%%%%%%%%%%%%%%%%%%%%%%%%%%%%%%%%%%%%%%%%
\conflictsofinterest{%{Declare conflicts of interest} % or state %``
The authors declare no conflict of interest.}%'' %Authors must identify and declare any personal circumstances or interest that may be perceived as inappropriately influencing the representation or interpretation of reported research results. Any role of the funders in the design of the study; in the collection, analyses or interpretation of data; in the writing of the manuscript, or in the decision to publish the results must be declared in this section. If there is no role, please state ``The funders had no role in the design of the study; in the collection, analyses, or interpretation of data; in the writing of the manuscript, or in the decision to publish the results''.} 


%=====================================
% References, variant A: internal bibliography
%=====================================
\reftitle{References}
%\begin{thebibliography}{999}
% Reference 1
%\bibitem[Author1(year)]{ref-journal}
%Author1, T. The title of the cited article. {\em Journal %Abbreviation} {\bf 2008}, {\em 10}, 142--149.
% Reference 2
%\bibitem[Author2(year)]{ref-book}
%Author2, L. The title of the cited contribution. In {\em The %Book Title}; Editor1, F., Editor2, A., Eds.; Publishing House: %City, Country, 2007; pp. 32--58.
%\end{thebibliography}
% The following MDPI journals use author-date citation: Arts, Econometrics, Economies, Genealogy, Humanities, IJFS, JRFM, Laws, Religions, Risks, Social Sciences. For those journals, please follow the formatting guidelines on http://www.mdpi.com/authors/references
% To cite two works by the same author: \citeauthor{ref-journal-1a} (\citeyear{ref-journal-1a}, \citeyear{ref-journal-1b}). This produces: Whittaker (1967, 1975)
% To cite two works by the same author with specific pages: \citeauthor{ref-journal-3a} (\citeyear{ref-journal-3a}, p. 328; \citeyear{ref-journal-3b}, p.475). This produces: Wong (1999, p. 328; 2000, p. 475)

%=====================================
% References, variant B: external bibliography
%=====================================
\externalbibliography{yes}
\bibliography{bibtex}

%%%%%%%%%%%%%%%%%%%%%%%%%%%%%%%%%%%%%%%%%%
%% optional

%% for journal Sci
%\reviewreports{\\
%Reviewer 1 comments and authors’ response\\
%Reviewer 2 comments and authors’ response\\
%Reviewer 3 comments and authors’ response
%}

%%%%%%%%%%%%%%%%%%%%%%%%%%%%%%%%%%%%%%%%%%

\newpage
\appendix
\section{Extended Methods} \label{extended_methods}

\subsection*{General remark}

At multiple instances, a gradual input needs to be translated into a clear on/off output around some boundary value, while ensuring a solvable ODE system. To elegantly solve for this, a sigmoid function is used, defined by $\sigma_k(x)=\frac{1}{1+\exp^{-k*x}}$, where $k$ indicates the function's steepness. The sigmoid is represented simply as $\sigma(x)$ wherever the standard steepness of $k=1$ was used. 

\subsection{Adapted SIR}
We use the classical SIR ordinary differential equation formulation without births or deaths~\cite{shulgin1998pulse} to calculate the number of susceptible $U$, infected $I$ and recovered $R$ individuals. (The regular expression for susceptibles $S$ is already used to refer to `stock' in this paper.) Additionally, we introduce the possibility of a delay in epidemic onset, by multiplying the differential equations by the sigmoid function $\sigma(t-\Delta t)$. $\Delta t$ indicates the number of days between $t=0$ and the start of an epidemic with $I(\Delta t)=1$. 

\begin{align} \label{SIR_ODEs}
\begin{split}
    \frac{dU}{dt} &= \sigma(t-\Delta t) * - \beta  U I  \\
    \frac{dI}{dt} &= \sigma(t-\Delta t) * \Big(\beta U I - \gamma I\Big) \\
    \frac{dR}{dt} &= \sigma(t-\Delta t) * \gamma I \\
\end{split}
\end{align}

% \begin{align} \label{SIR_ODEs}
% \begin{split}
%     \frac{dS(t)}{dt} &= \sigma(t-\Delta t) * - \beta (t) S(t) I(t)  \\
%     \frac{dI(t)}{dt} &= \sigma(t-\Delta t) * \Big(\beta (t) S(t) I(t) - \gamma I(t)\Big) \\
%     \frac{dR(t)}{dt} &= \sigma(t-\Delta t) * \gamma I(t) \\
% \end{split}
% \end{align}

Here $\beta$ indicates the infection rate and $\gamma$ the recovery rate. $\beta$ is calculated from the effective reproduction number $R_e$. This is defined as the ratio $R_e=\beta/\gamma$ and is a measure of the infectiousness of the virus: the average amount of people getting infected from one infected individual. It is assumed that the availability of PPE lowers the virus' infectiousness. Therefore the effective reproduction number is calculated from the basic reproduction number $R_0$ (infection spread without any protective measures).

\begin{equation} 
\label{Eq_R}
    R_{e}(t) =  R_0\Big(1-\sigma(S(t))r\Big)
\end{equation} 

with $S(t)$ the stock at time $t$ and $r$ the reduction in infectiousness if the stock is non-zero. It is thus assumed that $R_e$ is reduced to some fixed lower value depending only on $r$ and $R_0$ as long as $S>0$, but reverts back to the original $R_0$ value as soon as the stock has depleted ($S=0$).

\subsection{DSP model}

The DSP chain was modelled via the following differential equations for demand $D$, available stock $S$ and production $P$:

\begin{align}
  &\begin{aligned} \label{ODE_demand}
    \frac{dD(t)}{dt} = f(t) = wI(t)
  \end{aligned} \\
  &\begin{aligned} \label{ODE_stock}
    \frac{dS(t)}{dt} = P(t)-D(t)*\sigma(S(t))
  \end{aligned}\\
  &\begin{aligned}
    \frac{dP(t)}{dt} =   \Bigg(\frac{dD(t)}{dt}+&\Big(P^{\textrm{max}}-P(t)\Big)\Big(1-\frac{S(t)}{S^{\textrm{max}}}\Big) \sigma \Big(S^{\textrm{max}} - S(t)\Big)- \Big(P(t)-D(t)\Big) \sigma \Big(P(t)-D(t)\Big)\Bigg) \\
    & * \sigma \Big(P^{\textrm{max}}-P(t)\Big)
  \end{aligned}
\end{align}

Here $w$ is the PPE needed per infected individual. The demand thus grows linearly with the amount of infected. Interestingly, this demand equation can be replaced by any positive and continuous demand equation $f(t)$ of interest and could thus be adapted to be reused in other supply chain contexts. It is assumed that there is a baseline PPE demand $D(0)$, for regular non-epidemic related use. The baseline production $P(0)$ is matched to this standard demand ($D(0)=P(0)$). Before epidemic onset, the stock $S(t)$ is thus kept at a dynamic equilibrium, which we assume to be at its full capacity $S^{\textrm{max}}$. The stock difference is calculated from the difference between production and demand, but demand can only be subtracted if there is still stock available to prevent a negative stock. When the demand changes due to differences in the amount of infected, the production difference follows the difference in demand with a cap of the maximum production per day ($P^{\textrm{max}}$). Again, sigmoid functions are used to control for different scenario's in three ways: (1) There can only be an increase in production when the current production is below maximum production; (2) When the stock is below $S^{\textrm{max}}$, the production is increased further than the demand to enable stock replenishment; (3) When production is higher than demand, an additional force is added to prevent a stock overshoot, that gradually pushes the production back to the demand level.

We can see how the coupling between the SIR and DSP model parts is thus realised via two factors: (1) the demand equation~\ref{ODE_demand} that is dependent on $I(t)$ and (2) the $R_e$ calculation~\ref{Eq_R} with the current $S(t)$.

\subsection{Sharing mechanism}

The epidemic and production mechanism of each community \textit{i} can be separately described by the coupled SIR and DSP equations above. For product sharing between communities, a switch mechanism is constructed. As inputs, the current stock-maximum stock ratio's for both communities and the sharing threshold $\theta$ are taken. Sharing must be enabled when exactly one of the two communities' stocks drops below this critical threshold. The switch system is implemented as follows: 

\begin{equation} \label{eq:switch}
\begin{split}
  switch\Big(\frac{S_{A}}{S^{\textrm{max}}_A},\frac{S_{B}}{S^{\textrm{max}}_B},\theta\Big) = 
  2\times \Bigg[ 0.5 
  & - \sigma_k\Bigg(\sigma_k\Big(\frac{S_{A}}{S^{\textrm{max}}_A}-\theta\Big) + \sigma_k\Big(\frac{S_{B}}{S^{\textrm{max}}_B}-\theta\Big) - 2\Bigg) \\ 
  &- \sigma_k\Bigg(\sigma_k\Big(\theta-\frac{S_{A}}{S^{\textrm{max}}_A}\Big) + \sigma_k\Big(\theta-\frac{S_{B}}{S^{\textrm{max}}_B}\Big) - 2\Bigg)\Bigg]
\end{split}
\end{equation}

Due to the need for increased sensitivity, the steepness of the sigmoid was set higher than before, at $k=1,000$. The sharing mechanism was added to the separate community ODE's by adding shared stock ($S^{\textrm{share}}_{j \to i}$) to equation~\ref{ODE_stock} as follows:
\begin{equation} \label{ODE_stock_switch}
    \frac{dS_{i}}{dt}=\Big(P_{i}(t)-D_{i}(t)+ switch\Big(\frac{S_{A}}{S^{\textrm{max}}_A},\frac{S_{B}}{S^{\textrm{max}}_B},Th\Big)*S^{\textrm{share}}_{j \to i}\Big)\sigma(S_{i}(t) + \sigma(-S_{i}(t))
\end{equation}

Here the size of the shared PPE stock is calculated as $S^{\textrm{share}}_{j \to i}=(S^{\textrm{max}}_i + S^{\textrm{max}}_j)(\frac{S_i}{S^{\textrm{max}}_i}-\frac{S_{j}}{S^{\textrm{max}}_B})/2$. The amount of shared stock is thus related to the difference between the stocks of both communities. With the system described here we can simulate a whole spectrum of willingness to share, from a scenario with no sharing ($\theta=1.0$) to sharing the total stock ($\theta=0.0$). Mathematica~\cite{wolfram1999mathematica} was used to solve the system and plot the results. In Table~\ref{table:params} a brief outline is shown of the parameter settings used in the model.

%------------------------------------------
%\begin{wraptable}{r}{10cm}
\begin{table}
\centering

%\caption{\textbf{Parameter settings used for the simulations.}}\label{table:params}
\begin{tabular}{llrp{8cm}}
\textbf{Parameter}          & \textbf{Symbol} & \textbf{Value} & \textbf{Comment} \\
\hline
Population size             & $n$       & $1.7 \times 10^7$ & This parameter has no effect in the model but makes results more easily interpretable. It is roughly the population size of the Netherlands. \\
Basic reproduction number   & $R_0$     & $2.3$ & $R_0$ is the reproduction number of the virus in the absence of countermeasures and without immunity, i.e., at the start of the pandemic. In their June 2020 report\footnote{\url{https://www.who.int/docs/default-source/coronaviruse/risk-comms-updates/update-28-covid-19-what-we-know-may-2020.pdf}, accessed August 3, 2021.}, the WHO estimated this number in the range 2-4. In the Netherlands the peak estimate\footnote{\url{https://coronadashboard.rijksoverheid.nl/landelijk/reproductiegetal}, accessed August 5, 2021.} near the start of the pandemic is around $2.2$ but this varies significantly across countries and time. The most important about this parameter is that it should be larger than $1$ in order to lead to exponential growth and thus a population-wide epidemic.\\
% Infection rate              & $\beta$   & $R_e * \gamma$ &  \\
Recovery rate               & $\gamma$  & $1 / 6$  & This rate defines the average period ($1/\gamma$ days) during which an individual is infectious. In the same WHO report, the curve of virus shedding appears to be high for about six days. More recent sources report a \emph{median} of 8 days\footnote{\url{https://www.nature.com/articles/s41467-020-20568-4}, accessed August 5, 2021} of virus shedding. Together with the infection rate this parameter induces the reproduction number, which in our simulation is the more important number. \\
Initial production          & $P(0)$     & $5 \times 10^5$ & This parameter has an arbitrary unit which has no effect on the model but makes results more easily interpretable. According to a news article from April 2020, the Netherlands were using about $4.5 \times 10^6 / 7$ face masks per day, which is the most well-known resource. Our value here is thus chosen to be on the order of magnitude of the number of face masks. \\
% Initial demand              & $D(0)$    & $P(0)$ \\
Maximum production          & $P^{\textrm{max}}$ & $4 \times P(0)$ & A community can increase the local production to this quantity. This includes all sources, such as increasing local production as well as importing more resources from global suppliers. \\
Necessary PPE per infected case     & $w$   & $4$ & Keeping to face masks for interpretability, numbers vary wildly: from roughly 144 face masks per day per corona patient in the intensive care unit\footnote{\url{https://lc.nl/friesland/Coronapati\%C3\%ABnt-op-ic-vergt-144-mondmaskers-per-24-uur-25488723.html}, accessed August 2021.} to one or zero for infected persons who are either unaware of their infection or remain in quarantine at home. For interpretability we turn to what would in principle be needed to prevent one infected person from infecting others as much as possible. According to various sources a disposable mask should only be used for about three hours. This would mean that in principle in order to cover one's face for 12 hours per day, about 4 masks would be needed, which is the (arbitrary) quantity we chose here. \\
Reduction in $R_e$ for $S>0$ & $r$   & $0.4$ & This factor multiplies with the infection rate and thus reduces the reproduction number. Its value is arbitrary with the most important property being that $r \cdot R_0 < 1$. In words this means that having sufficient PPE would stop the exponential growth of the epidemic. \\
% Maximum stock               & $S^{\textrm{max}}$    & $\in [1 \times 10^7, 7 \times 10^7]$ \\
% Sharing threshold           & $\theta$  & $\in [0.0,1.0]$ \\
% Time difference epidemic onset  & $\Delta t$    & $\in [-200, 200]$ \\

\end{tabular}
\caption{\textbf{Parameter settings used for the simulations in this paper.}}
\label{table:params}
\end{table}
%\end{wraptable} 
%------------------------------------------

%{\lipsum[2] 
%\par
%Table~\ref{table:params} is a wrapped table.
%\\
%\\
%\textcolor{red}{Could still include: \\
%- initial settings; \\
%- ranges of the inputs we vary (maximum stock, sharing threshold, $\Delta t$); \\
%- units. But many of our params don't have obvious units.}

\newpage
% \section{Supplementary Figures} \label{supp figures}
% \beginsupplement

\begin{figure}[b!]
    \centering
    \includegraphics[width=0.8\textwidth]{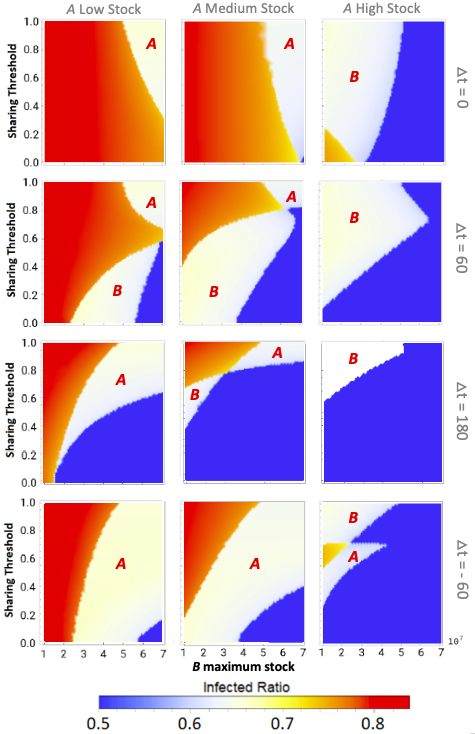}
    \caption{\textbf{Influence of $B$ maximum stock and sharing threshold on total infected ratio for different epidemic onset time differences.} The epidemic is introduced in $A$ at $t=0$, and in $B$ at $t=\Delta t$. The maximum stock of $A$ is fixed at either Low Stock ($S^{\textrm{max}}=1\times10^7$), Medium Stock ($S^{\textrm{max}}=3\times10^7$) or High Stock ($S^{\textrm{max}}=7\times10^7$). $n=1.7\times10^7$, $R_{0}=2.3$, $\gamma=1/6$, $r=0.4$, $w=4$, $P(0)=5\times10^7$, $P^{\textrm{max}}=4\times P_i(0)$, $D(0)=P(0)$.}
    \label{fig:diffMaxStock_C2Init}
\end{figure}

\newpage

\end{document}